\documentclass[useAMS,usenatbib]{mn2e}
 
\def\chandra{{\it Chandra}}

\usepackage[dvips]{graphicx}
\usepackage{graphics}
\usepackage{lscape}
\usepackage{amssymb}
\usepackage{mathrsfs}
\input{epsf}

\title[On the X-ray properties of submm--selected galaxies]
{On the X-ray properties of submm--selected galaxies}

\author[E. S. Laird et al.]
{Elise S. Laird$^{1}$\thanks{E-mail:~e.laird@imperial.ac.uk}
Kirpal Nandra$^{1}$, Alexandra Pope$^{2}$, Douglas Scott$^{3}$ \\
$^{1}$Astrophysics Group, Imperial College London, Blackett Laboratory,
Prince Consort Road, London SW7 2AZ, UK\\
$^{2}$National Optical Astronomy Observatory, Tucson, AZ 85719, USA \\
$^{3}$Departmanet of Physics and Astronomy, University of British Columbia, Vancouver, BC V6T 1Z1, Canada}
\begin{document}

\date{Accepted 0000 December 00. Received 0000 December 00; in original form 0000 October 00}

\pagerange{\pageref{firstpage}--\pageref{lastpage}} \pubyear{2009}

\maketitle

\label{firstpage}
\begin{abstract} 
We present an analysis of the X-ray properties of 35 submm galaxies (SMGs) in
the {\it Chandra\/} Deep Field North region.  Using a sample of robust
$850\,\mu$m--selected galaxies, with sub-arcsecond positions from {\it Spitzer\/}
and/or radio counterparts, we find 16 objects ($45\pm8$ per cent) with
significant X-ray detections in the 2-Ms {\it Chandra\/} data.
6 of these SMGs ($\sim17\pm6$ per cent) have measured X-ray luminosities or upper limits consistent
with those expected based on the far-infrared or radio-derived star formation rate, and
hence with the X-rays coming solely from star-forming processes. Extrapolating observed X-ray/star formation rate relations to the luminosity of the SMGs we find that the X-ray derived star formation rates are 
typically in the range 200--$2000\,{\rm M}_{\odot}\,{\rm yr}^{-1}$. In another 7 sources ($20\pm7$ per cent of the SMGs)
a dominant AGN contribution to the X-ray emission is required, while in 3 
more it is unclear whether stellar process or accretion are responsible for the X-rays.
Stacking of the X-ray undetected SMGs reveals a highly significant (7$\sigma$)
detection. Under the assumption that the stacked X-ray is due to star formation this gives an 
average X-ray derived star formation rate of around $150\,{\rm M}_{\odot}\,{\rm yr}^{-1}$.  
We deduce that the AGN fraction in SMGs based on X-ray observations is $20-29\pm7$~per cent, which is 
towards the lower limit of
previous estimates.  Spectral analysis shows that in general the SMGs are not
heavily obscured in the X-ray but most of the SMGs classfied as AGN show absorption with
column densities in excess of $10^{22}\,{\rm cm}^{-2}$.  Of the secure AGN, the
bolometric luminosity appears to be dominated by the AGN in only 3 cases.  In
around 85~per cent of the SMGs, the X-ray spectrum effectively rules out an AGN
contribution that dominates the bolometric emission, even if the AGN is Compton
thick.  The evidence therefore suggests that intense star formation accounts
for both the far-infrared and X-ray emission in most SMGs.  We argue that, rather than
having an especially high AGN fraction or duty cycle, SMGs have a high X-ray
detection rate at very faint fluxes partly because of their high star formation
rates and, in rarer cases, because the submm emission is from an AGN.
\end{abstract}

\begin{keywords}
galaxies: active -- galaxies: starburst -- galaxies: high-redshift --
 X-rays: galaxies
\end{keywords}

\section{Introduction}


\begin{table*}
\begin{minipage}{180mm}
\begin{center}
\caption{X-ray detected submm sources.
Col.(1): Submm source name \citep{pope05,wall08}.
Col.(2): Redshift.  The bold faced redshifts are spectroscopic, otherwise they
are photometric.  Redshifts in italic are from IRS spectroscopy.  Redshifts in
parentheses are estimated from IRAC colours, for those sources without optical counterparts. Optical photometric redshifts have {\tt ODDS} parameter $>0.9$ \citep{pope05}. IRAC photometric redshifts have an accuracy of $\sigma[\Delta z/(1+z)]\sim0.1$).
Col.(3): $i_{775}$  magnitude from {\it HST}/ACS imaging.
Col.(4): Flux-deboosted $850\,\mu$m flux density.
Col.(5): IR (8-1000$\mu$m) luminosity. $L_{\rm IR}$ is calculated from the rest frame
$L_{\rm 160}$ by scaling by the typical $L_{\rm 160}$-to-$L_{\rm IR}$ conversion from Table A2 of \citet{pope06}.
Col.(6): Radio luminosity.
Col.(7): {\it Chandra\/} name
Col.(8): {\it Chandra\/} RA.
Col.(9): {\it Chandra\/} Dec.
Col.(10): Positional offset between IRAC and {\it Chandra\/} position (arcsec), after correcting for a small (0.6 arcsec)
overall offset  between the two coordinate frames. 
X-ray data are from  \citet{nandra05} and \citet{laird05}, all other data from
\citet{pope05,pope06,pope08}, \citet{wall08} and references therein. The redshift for GN10 is from \citet{daddi09}.
}
\label{sample}
\begin{tabular}{@{}lccccccccc@{}}
\hline
Name  &  $z$ & $i_{775}$ & $S_{\rm 850}$ & $L_{\rm IR}$ & $\log L_{\rm 1.4 GHz}$ & CXO & RA & DEC & Off  \\
      &   & (AB)  & (mJy) & ($10^{12}$L$_{\odot}$) & (W Hz$^{-1}$) & HDFN & (J2000) & (J2000) & (\arcsec)  \\                
(1)  & (2) & (3) & (4)               & (5)         & (6)            & (7) & (8) & (9) & (10)  \\ 
\hline
GN01$^{*}$  &\bf{2.415} & 23.3 & $6.2 \pm 1.6$& 7.38  & 23.95 &J123606.6+621550 & 189.027800 & 62.263960& 0.337   \\
GN04$^{a,*}$  &\bf{2.578} & 26.2 & $4.9\pm0.7$  & 5.86  & 24.54 &J123616.0+621513 & 189.066940 & 62.253730& 0.426   \\
~~~~~~~$^{b}$  &                  & $>$28&      &        &     &J123615.7+621515 & 189.065701 & 62.254219& 0.402  \\
GN06$^{*}$  & {\bfseries{\emph{2.00}}}& 27.4 & $7.5\pm 0.9$ & 8.73  & 24.59 &J123618.3+621550 & 189.076550 & 62.264060& 0.874  \\
GN10  &\bf{4.042}   & $>$28& $11.3\pm1.6$ & 15.10 & 24.54  &   J123633.3+621408 & 189.139150 & 62.235570 & 0.226   \\
GN12  &3.1   & 26.2 & $8.6\pm 1.4$ & 10.68 & 24.78      & J123646.0+621448 & 189.191770 & 62.246770 & 0.175   \\
GN13  &\bf{0.475} & 21.6 & $1.9\pm 0.4$ & 0.94  & 22.54 &J123649.6+621312 & 189.206970 & 62.220130& 0.165  \\
GN15  &\bf{2.743} & 24.3 & $3.7\pm 0.4$ & 4.48  & ...   &J123655.7+621200 & 189.232470 & 62.200150& 0.079  \\
GN17  &{\bfseries{\emph{1.73}}}  & 27.7 & $3.9\pm 0.7$ & 4.41  & 24.23 &J123701.5+621145 & 189.256610 & 62.196010 & 0.195 \\
GN19$^{*}$  &\bf{2.484} & 25.4 & $8.0\pm 3.1$ & 9.60  & 24.44 &J123707.1+621407 & 189.279970 & 62.235370& 0.246   \\
GN22$^{*}$  &\bf{2.509} & 24.6 & $10.5\pm4.3$ & 12.64 & 24.38 &J123606.8+621021 & 189.028380 & 62.172510& 0.187  \\
GN23  &(2.6)   & $>$28& $4.9\pm 2.3$ & 5.97  & 24.14     &J123608.5+621434 & 189.035630 & 62.243010 & 0.157   \\
GN25$^{*}$  &\bf{1.013} & 22.8 & $3.2\pm 1.4$ & 2.98  & 23.63 &J123629.0+621045 & 189.121180 & 62.179280& 0.066   \\
GN26$^{*}$  &\bf{1.219} & 22.7 & $2.2\pm 0.8$ & 2.23  & 24.13 &J123634.4+621240 & 189.143590 & 62.211280& 0.240   \\
GN30  &\bf{1.355} & 22.7 & $1.8\pm 0.5$ & 1.87  & 23.25 &J123652.7+621353 & 189.219820 & 62.231660& 0.260   \\
GN39$^{b}$  &\bf{1.996} & 23.4 & $5.2\pm 2.4$ & 6.05  & 24.63 &J123711.3+621331 & 189.297150 & 62.225122 & 0.121  \\
~~~~~~~~$^{a,*}$  & \bf{1.992}         & 24.9 &		   &  	   &             &J123712.0+621325 & 189.299835 & 62.223630 & 0.429  \\
GN40  &(2.6)      &$>$28  &$10.7\pm 2.9$ & 12.92 & 25.39     &J123713.8+621826 & 189.307681 & 62.307152 & 0.074  \\
\hline
\end{tabular}
\end{center}
$^*$Included in the A05b sample. The redshift used for GN06 in A05b was 1.865.\\
$^a$Southern X-ray and IR counterpart.\\
$^b$Northern X-ray and IR counterpart.\\
\end{minipage}
\end{table*}


There has been vigorous debate about whether the most luminous galaxies, most of which have been discovered
through their far-infrared (FIR) emission (e.g.~\citealt{sanders96,rowan00})
are violent starbursts, or obscured active galactic nuclei (AGN), with the consensus appearing to be that both processes are important
(e.g.~\citealt{veilleux95,farrah02}).  This fits in with the idea of a
`starburst-AGN connection' where accretion and star formation, both plausibly
triggered by mergers, occur coevally and play a complementary role in the
formation of galaxies (e.g.~\citealt{springel01,hopkins05}).  

Highly luminous galaxies selected by their submm-emission have attracted particular attention. This is in large part due to the great observational progress afforded by the SCUBA instrument on the JCMT \citep{holland99}, which has vastly increased the number of known SMGs. While it is clear that submm-selected sources represent a diverse population (e.g.~\citealt{ivison00}), their nature has been the subject of intense study. These efforts stem from two key observations: firstly that if SMGs are dominated by star formation they are among the most powerful starbursts -- and most luminous sources -- in the Universe (e.g. \citealt{smail97,barger98,hughes98,lilly99}). Secondly, as these galaxies are reasonably numerous, they may contribute a significant, or even dominant contribution to the global star formation rate (e.g.~\citealt{blain99,chapman05,aretxaga07,wall08}). Additionally, whether or not SMGs are dominated by star formation or AGN processes, it seems clear that this sub-mm bright phase is a crucial one in the evolution and formation of the most massive galaxies. 

Pivotal developments in our understanding of the submm population have recently been achieved thanks to deep optical and mid-infrared spectroscopic, and X-ray observations. \citet{chapman03,chapman05} have identified a large number of SMGs associated with optically faint radio sources  in several different fields, following up the counterparts with deep Keck spectroscopy. These observations, in addition to providing the first significant sample of secure redshifts for SMGs, have shown that they occur at a median redshift $z \sim 2$. Recently, a few rare SMGs have been found out to $z\sim4-5$ \citep{capak08,daddi09} but the bulk of this population is still consistent with a median redshift of 2--3. This contrasts with some earlier predictions that the SMG population might be at extremely high redshifts. 

Using the \citet{chapman05} source list of SMGs, Alexander at al. (2005a, hereafter A05a; 2005b hereafter A05b) have identified X-ray counterparts to a sample of 20 SMGs in the {\it Chandra\/}-Deep Field North (CDF-N) region, where ultradeep X-ray data are available. The \citet{chapman05} sample of SMGs that have spectroscopically identified, optically-faint radio counterparts is representative of approximately half of all SMGs. Based on the analysis of this sample A05a argued that the AGN fraction in SMGs is very high ($>38^{+12}_{-10}$ per cent), and hence that the intense star formation in SMGs is accompanied by coeval and near-continuous black hole growth. A05b presented a detailed spectral analysis of the X-ray detected submm galaxies, concluding that $75$ per cent of radio-selected, spectroscopically identified SMGs host AGN, and that the majority ($\sim 80$ per cent) of these are highly obscured, with column densities $>10^{23}$~cm$^{-2}$. Despite the apparently high AGN fraction, mid-IR spectroscopy of SMGs suggests that the AGN is generally not the dominant contributor to the bolometric luminosity \citep{valiante07,pope08,menendez09}, so the conclusion that they are vigorously forming stars still holds. 

Overall, the above results argue that many of the most massive galaxies in the Universe undergo a period of intense star formation activity and black hole growth at moderate-to-high redshifts, which is deeply shrouded in gas and dust. This coeval mode of activity could plausibly result in the local correlations seen between black hole masses and bulge properties \citep{magorrian98,gebhardt00,ferrarese00}. A note of caution should be sounded, however, as the \citet{chapman03,chapman05} method of identifying SMGs, while apparently highly successful, could potentially suffer selection biases due to the requirement of radio emission, optical faintness and spectroscopic redshifts. 

Pope et al. (2006; hereafter P06), have used a new reduction of the radio data in the CDF-N together with deep {\it Spitzer\/}
observations from the GOODS project to identify secure counterparts to a large number of high significance SCUBA $850\,\mu$m sources in the GOODS-N region of the CDF-N. The GOODS-N SCUBA survey is small and non-uniform but to date it remains the submm field with the most complete counterpart identification and deepest X-ray data.  In this paper, we use the P06 sample to re-examine the X-ray emission of SMGs and in particular to assess the AGN fraction and X-ray spectral properties. 

Throughout, a standard, flat $\Lambda$CDM cosmology with $\Omega_\Lambda = 0.7$ and H$_0 = 72$~km~s$^{-1}$~Mpc$^{-1}$ is assumed.

\section{Observations and data reduction}

\subsection{The {\it Spitzer}/submm sample}

The submm galaxies used in the work have been selected from the SCUBA Super-map of the GOODS-N region of the CDF-N. The SCUBA Super-map is composed of data observed in all SCUBA modes (photometry, jiggle mapping and scan mapping) and thus provides inhomogeneous coverage and depth across the GOODS-N field (see \citealt{borys03,pope05} and P06 for a full description of the Super-map and corresponding analysis). Taking into account the varying depth across the field P06 have presented a sample of 35 $850\,\mu$m selected sources in the GOODS-N region, and using radio and {\it Spitzer\/} data have identified highly reliable counterparts to 21 of these, with likely counterparts to an additional 12. An update and re-analysis of the SCUBA Super-map with additional observation in the fields resulted in two further $850\,\mu$m selected sources with highly reliable counterparts, and is described in the appendix of \citet{wall08}. 

For this sample of 35 identified submm galaxies we therefore have accurate radio or {\it Spitzer\/} positions, which we can then use to unambiguously associate with X-ray counterparts. P06 have identified the optical counterparts to the SMGs where they are detected in the {\it HST\/} images \citep{giavalisco04}. They also assembled together all available redshift information, including the available spectroscopic redshifts, photometric redshifts for objects detected optically, and {\it Spitzer}-derived photometric redshifts for optically blank SMGs. There are three sources with only photometric redshifts and while there is some uncertainty in these, particularly in the {\it Spitzer}-derived redshifts, for our purposes small redshift errors of order $\Delta z$~$\sim$few tenths will have little effect on our conclusions. It should be noted that GN10 has recently been identified via CO line emission as being at $z=4.042$ \citep{daddi09}, considerably higher than the previous photometric redshift estimate of $z=2.3$, and we use the updated redshift value here. Nine of the SMGs in this sample have also been observed with the {\it Spitzer\/} IRS by \citet{pope08} to confirm their redshifts and determine the level of AGN contribution.

The basic properties of the SMGs with X-ray detections are given in Table~\ref{sample}.

\subsection{X-ray data and reduction}

In this paper we use the analysis of the 2-Ms {\it Chandra\/} ACIS-I observation of the CDF-N, which is the deepest X-ray exposure yet taken \citep{alexander03}. The analysis is described in detail in \citet{laird05}. 

Briefly, after applying relatively standard screening procedures we created images and exposure maps  in several bands, notably the full band (0.5--$7\,$keV), hard band (2--$7\,$keV) and soft band (0.5--$2\,$keV). The source detection algorithm described in \citet{nandra05} was then applied, which yields a similar, but slightly larger number of sources than used by \citet{alexander03}. The positions of X-ray sources in this catalogue were then correlated with those of the submm sources from P06. The positional offsets between the X-ray sources and the submm sources, after the removal of a small overall offset, are given in Table~\ref{sample}.

Spectra were extracted for all submm/X-ray associated sources using the methodology described in \citet{laird06}. {\small CIAO} v3.2.1 and {\small CALDB} v3.0.1 were used for the spectral extraction and {\small XSPEC} v11.3.1 software used for the spectral analysis. The energy range of all of the analysis was restricted to 0.5--$7\,$keV.
Because the  CDF-N dataset has been taken in several different observing configurations and positions, spectra must be extracted separately from each observation, and we have then summed these together using the standard {\small FTOOLS} \citep{blackburn95} routines. Source counts were extracted from a circular region equal to the 95 per cent encircled energy fraction (EEF) of the \chandra\ point-spread function (PSF) at $2.5\,$keV. Local background regions were manually selected to avoid contamination by nearby X-ray sources. Typically, the background was extracted from three to four large regions surrounding the source. The effective area and response files were then combined, weighted by exposure, using the {\small FTOOLS} routines {\small ADDARF} and {\small ADDRMF}. As discussed by \citet{laird06} the first three \chandra\ observations were taken at a different focal plane temperature to the remainder of the observation. We therefore restricted our analysis to the 17 observations taken at a focal plane temperature of $-120$\degr C, reducing the total exposure of the spectra by 161.7~ks. 

Because of the small number of counts in the individual spectra the spectral fitting was performed using a maximum likelihood method (C-statistic; \citealt{cash79}). This has the disadvantage that the goodness-of--fit cannot be easily assessed (but see \citealt{lucy00}). It is possible, however, to use the C-statistic to assess whether a particular model component is needed in a particular dataset and, more importantly, to assign confidence intervals to the parameters. All of the spectra analysed using the C-statistic were grouped into fixed width bins of 4 channels ($\sim700$~eV), which greatly improves the processing times in {\small XSPEC} without significant loss of information for these photon-starved spectra.  

For GN04 and GN39, we identified two potential X-ray counterparts (see below).  Spectra were extracted from each of the individual X-ray counterparts that are associated with the IRAC components of the sub-mm galaxies. For GN04, where the X-ray sources are only separated by 2.5 arcsec and therefore are not fully resolved by the {\it Chandra\/} PSF we extracted the spectra using a region smaller than the 95 per cent to minimise contamination by the neighbouring source. 

\begin{figure}
\begin{center}
\includegraphics[width=75mm,angle=90]{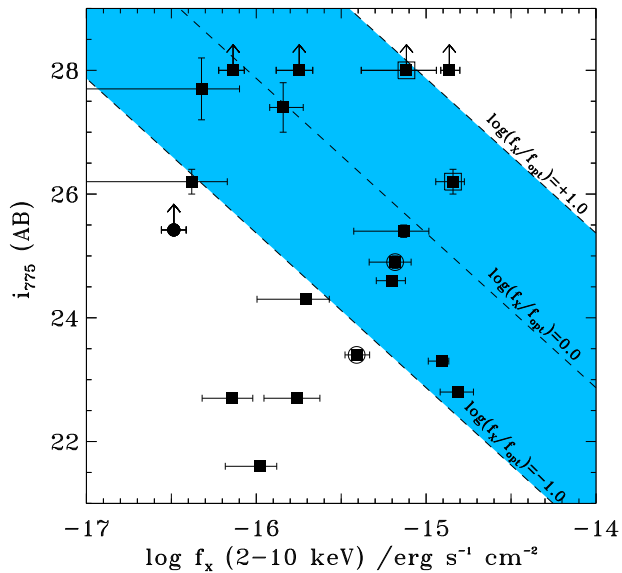}
\caption{2--$10\,$keV X-ray flux vs. $i_{775}$ magnitude (AB) for the X-ray
detected 
submm galaxies (filled squares). Fluxes are calculated for the spectral fits using Model A (see section 3.2). Both of the optical/IR and X-ray counterparts to the SMG are plotted for GN04 and GN39, and are identified with open squares and circles, respectively. The filled circle shows the 2--$10\,$keV stacking result for the undetected SMGs. Not all of the SMGs in the stacked sample are detected in the optical, therefore the mean  $i_{775}$ magnitude for the stacked sample is shown as a lower limit.
Diagonal lines indicate constant X-ray to optical flux ratios. The $i_{775}$ magnitudes were converted using $0$~mag$=1.405\times10^{-7} \rmn{erg~s}^{-1}\rmn{cm}^{-2}$.  
}
\label{fxfopt}
\end{center}
\end{figure} 


\section{RESULTS}

\subsection{X-ray and submm associations}

The most basic information is the fraction of submm sources associated with X-ray sources. Cross-correlation of the IRAC positions of the \citet{pope06} SMGs with the \citet{nandra05} X-ray source list gives a total of 16 matches within 1 arcsec. The probability that any of these are a random association is 0.13 per cent. There are 4 sources in the SMGs catalogue which have two IRAC/radio sources in the SCUBA error circle, both of which may contribute to the submm flux. Of these GN04, GN19 and GN39 have X-ray counterparts. In the case of GN19, the X-ray counterpart is clearly identified with the first counterpart listed by \citet{pope06}, i.e. the optically brighter source. In the cases of GN04 and GN39 there are X-ray counterparts for both of the radio sources.  We therefore find an X-ray source detection fraction for submm-selected galaxies of $46\pm 8$~per cent, where the error bars are calculated for a binomial distribution.

All the detected submm sources are extremely weak in the X-ray, with 2--$10\,$keV fluxes ranging from ranging from $4 \times 10^{-17}$~erg cm$^{-2}$ s$^{-1}$ (close to the limit of the survey) to $2 \times 10^{-15}$~erg cm$^{-2}$ s$^{-1}$, still a factor $\sim 6$ below the break in the X-ray number counts (e.g.~\citealt{kim07,georgakakis08}) where most of the X-ray background is produced. The X-ray-to-optical flux ratios of the detected sources covers a wide range, with many showing properties in the ``normal'' range (expected for X-ray selected AGN), others relatively optically bright and two which are undetected even in the deep {\it HST\/} imaging ($I_{\rm AB}>28$).  The $F_{\rm X}/F_{\rm opt}$ diagram (Fig. \ref{fxfopt}) is often used as a diagnostic tool to help in source classification (e.g.~\citealt{schmidt98,giacconi01,horn03}). In this case, given the extreme faintness of the sources in the X-ray, their wide redshift distribution, and the likelihood that many of the objects are heavily obscured by dust, these classifications are likely to have limited meaning. It is, however, clearly evident that SMGs are a heterogeneous class in terms of their X-ray to optical flux ratio. 

\subsection{X-ray Spectral fitting}.

\begin{table*}
\begin{minipage}{180mm}
\begin{center}
\caption{X-ray spectral properties of submm galaxies. All errors correspond to the 90 per cent confidence level. 
Col.(1): Submm source name.
Col.(2): Total (source+background) counts in the 0.5--$8\,$keV band.
Col.(3): Estimated background counts in source extraction region. 
Col.(4): Hardness ratio (H-S)/(H+S) where H and S are the count rates in the HB
and SB, corrected to on-axis values. 
Col.(5): Exposure-weighted off-axis angle.
Col.(6): Effective photon index for a fit with only Galactic absorption. Errors
Col.(7): Inferred intrinsic $N_{\rm H}$ from fits to Model A
{\small (WABS*ZWABS*(POWERLAW+PEXMON))} with fixed $\Gamma$=1.9 and pexmon normalisation
fixed to powerlaw normalisation).
Col.(8): Improvement in the C-statistic for Model A compared to a $\Gamma$=1.9
power-law spectrum with no intrinsic absorption (Model C).
Col.(9): Observed frame 2--$10\,$keV flux.
Col.(10): Log rest-frame 2--$10\,$keV luminosity (observed).
Col.(11):Log rest-frame 2--$10\,$keV luminosity (absorption corrected).
Col.(12): X-ray derived SFR, using the \citet{ranalli03} relation.
Col.(13): Source classification based on X-ray properties. `Amb' denotes sources where the X-ray emission could be from an AGN or star formation. 
}
\label{fits}
\begin{tabular}{@{}lcrccccccccccc@{}}
\hline
Name  & S & B & HR & OAA  & $\Gamma$ & $N_{\rm H}$ & $\Delta$C-stat &$F_{\rm X}$ & $L_{\rm X, uncor}$ & $L_{\rm X, cor}$ & SFR$_{\rm X}$ &Class\\
      & &   &  & arcmin & &  $10^{22}$~cm$^{-2}$  & & erg cm$^{2}$ s$^{-1}$   & erg s$^{-1}$  & erg s$^{-1}$ & M$_{\sun}~\rmn{yr}^{-1}$\\                
(1)  & (2) & (3) & (4)               & (5)         & (6)            & (7) & (8) & (9) &(10) &(11) &(12) &(13) \\ 
\hline
GN01 &373 & 95 & $-0.58$& 5.1 & 2.01$^{+ 0.22}_{- 0.25}$ & $<$2.2		     & 0.0   & 1.24$\times10^{-15}$ & 43.62 &       &      & AGN\\
GN04$^{a}$  &98 & 8 & 0.65  & 3.8 &$-0.32^{+ 0.33}_{- 0.35}$ & 107.1$^{+30.7}_{-26.3}$ & 121.9 & 1.44$\times10^{-15}$ & 42.80 & 43.92 & & AGN       \\
~~~~~~~~$^{b}$  &125 & 10 & 0.06  & 3.8 & 0.95$^{+ 0.28}_{- 0.26}$ & 15.8$^{+7.2}_{-5.5}$    & 30.7  & 7.66$\times10^{-16}$ & 43.18 & 43.52 &      &  \\
GN06 &105 & 58 & $-0.26$& 3.8 & 1.47$^{+ 0.87}_{- 0.69}$ & $<$7.4		     & 0.0   & 1.44$\times10^{-16}$ & 42.43 &       & 546  & SB\\
GN10 &34 & 19  & $-0.35$& 1.7 & 1.40$^{+ 1.46}_{- 1.36}$ & $<$70.8		     & 0.7   & 7.35$\times10^{-17}$ & 42.58 & 42.93 & 1702  & Amb\\
GN12 &25 & 20 & $-9.99$& 0.7 & 2.76$^{+ 2.36}_{- 1.35}$ & $<$20.2		     & 0.0   & 4.17$\times10^{-17}$ & 42.41 &       & 510  & SB\\
GN13 &44 & 18 & $-1.00$& 1.0 & 2.42$^{+ 1.37}_{- 0.90}$ & $<$0.5		     & 0.0   & 1.05$\times10^{-16}$ & 40.90 &       & 16   & SB\\
GN15 &71  & 23 & $-0.42$& 2.3 & 1.85$^{+ 0.55}_{- 0.59}$ & $<$8.9		     & 0.1   & 1.97$\times10^{-16}$ & 42.91 &       & 1625 & Amb\\
GN17 &31  & 24 & $-1.00$& 2.9 & 4.12$^{+ 4.02}_{- 1.93}$ & $<$3.3		     & 0.0   & 4.76$\times10^{-17}$ & 41.87 &       & 149  & SB \\
GN19 &92  & 21 & 0.16 & 2.3 & 0.58$^{+ 0.31}_{- 0.42}$ & 32.6$^{+15.2}_{-13.8}$   & 34.9  & 7.38$\times10^{-16}$ & 42.96 & 43.50 &      & AGN\\
GN22 &146 & 100 & 1.00 & 6.1 &$-0.33^{+ 0.82}_{- 0.98}$ & 81.0$^{+97.2}_{-42.2}$   & 15.1  & 6.29$\times10^{-16}$ & 42.58 & 43.51 &      & AGN\\
GN23 &119 & 71&  $-0.49$& 4.6 & 1.90$^{+ 1.16}_{- 0.76}$ & $<$4.6		     & 0.0   & 1.79$\times10^{-16}$ & 42.74 &       & 1106 & Amb\\
GN25 &202 & 35 & 0.31 & 4.0 & 0.50$^{+ 0.21}_{- 0.27}$ & 8.0$^{+2.6}_{-2.2}$	     & 82.8  & 1.54$\times10^{-15}$ & 42.66 & 42.89 &  & AGN \\
GN26 &65 & 23 & $-0.35$& 2.1 & 1.90$^{+ 0.60}_{- 0.63}$ & $<$2.1		     & 0.0   & 1.74$\times10^{-16}$ & 42.07 &       & 242  & SB\\
GN30 &32 & 17 & $-0.46$& 0.6 & 1.90$^{+ 2.11}_{- 0.98}$ & $<$3.9		     & 0.0   & 7.17$\times10^{-17}$ & 41.79 &       & 128  & SB\\
GN39$^{b}$  & 60 & 13 & 0.12 & 2.8 & 0.80$^{+ 0.46}_{- 0.51}$ & 9.40$^{+11.4}_{-5.9}$    & 8.7   & 3.20$\times10^{-16}$ & 42.63 & 42.88 &      & AGN\\
~~~~~~~$^{a}$      & 48 & 15 & 1.00 & 2.9 &$-0.29^{+ 0.27}_{- 0.92}$ & 67.1$^{+41.75}_{-23.2}$   & 51.0  & 7.07$\times10^{-16}$ & 42.52 & 43.34 &      & \\
GN40 &308 & 73 & $-0.27$& 5.3 & 1.24$^{+ 0.24}_{- 0.18}$ & 8.9$^{+3.6}_{-3.4}$      & 24.3  & 1.37$\times10^{-15}$ & 43.53 & 43.77 &      & AGN\\
\hline
\end{tabular}
\end{center}
$^a$Spectrum of Southern X-ray source.\\
$^b$Spectrum of Northern X-ray source.\\
\end{minipage}
\end{table*}

\begin{figure}
\begin{center}
\includegraphics[width=75mm,angle=90]{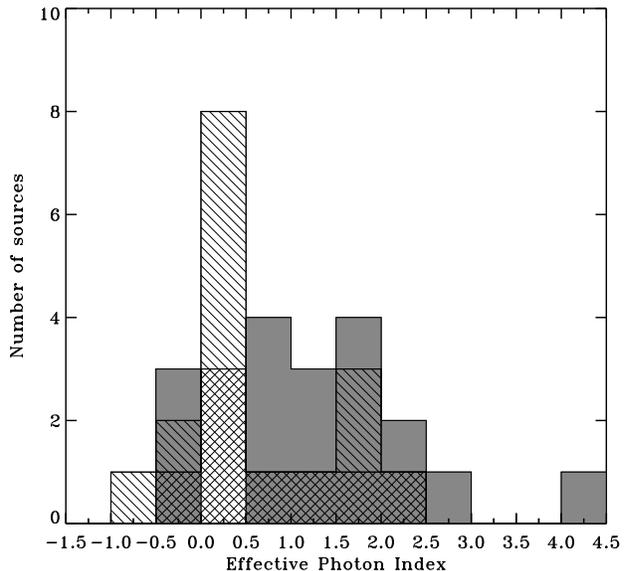}
\caption{Histogram of the effective photon-index of the SMGs, assuming no intrinsic absorption (see Table \ref{fits}). The distribution of photon-indices for this sample (grey shaded region) and the 17 sources with effective $\Gamma$ in Table 1 of A05b (hatched region) are shown. For GN04 and GN39 the photon-indices for both components are shown. The double hatched region shows the A05b photon-indices for the 8 sources common to both samples (as indicated in Table~\ref{sample}). Plotting the photon-indices for the common sources according to their values in Table \ref{fits} from this work shows a very similar, but not identical, distribution. Including (excluding) the 8 common sources, the K-S probability that the two samples are drawn from the same underlying galaxy distribution is 0.02 (0.005).}
\label{gamma_hist}
\end{center}
\end{figure} 

To proceed further in our analysis of these objects requires a consideration of their spectra. The work of A05b has indicated that as a class SMGs can be heavily obscured, and some may even be Compton thick. If so, even a quantity as simple as the luminosity requires a careful consideration of the effects of absorption and hence the spectral shape. In particular, at the highest column densities, the AGN may only revealed in scattered light, with the spectrum dominated by Compton reflection. Direct fitting of such models to the spectrum can thus constrain the contribution even of heavily buried AGN which might be rendered otherwise invisible. 

In the modelling of the spectra several different models are assumed. To characterise the spectra in the simplest manner all of the sources were first fitted by a Galactic absorption ($N_{\rm H}=1.5\times10^{20}$ cm$^{-2}$) plus power-law model ({\small WABS*POWERLAW} in {\small XSPEC}, Table~\ref{fits}, column 5) to determine the effective $\Gamma$. In order to characterise any intrinsic absorption we then fit the spectra with two different models.  The first (Model A) comprises of an intrinsically absorbed power-law representing emission from an AGN. This characterizes transmission dominated spectra. The second model (Model B) is a neutral reflection model with an unabsorbed power-law representing the direct component and a reflection component with the same photon index as the direct component. This model characterizes reflection-dominated spectra, which are often exhibited by Compton thick sources.

Given the very low number of counts in all the objects, this is extremely challenging, and this forces us to make a number of assumptions during the modeling of the spectra. In particular, for the majority of these low--count observations it is not possible to constrain simultaneously both the power-law photon index, $\Gamma$, and the line of sight absorbing column, $N_{\rm H}$.  For all models we therefore fixed the photon index in the power-law and reflection components at  $\Gamma=1.9$. The neutral reflection model is represented by the {\tt pexmon} model of Nandra et al. (2007), which includes both continuum reflection \citep{magdziarz95} and the associated Fe line emission \citep{george91}. For this model we assumed a single power law illuminating spectrum, and the abundances were fixed at the solar value. The assumed inclination was 60$^{\circ}$ and the solid angle of the reflector $\Omega/2\pi$ was fixed to 1. The strength of the reflection is therefore parameterized by $A_{\rm ref}$, the normalization of the reflection component, with the ratio of this to the power law normalization, $A_{\rm pl}$ representing the reflection fraction $R$ (see Nandra et al. 2007 for details). Note that the values of the reflection fraction, $R$, could be overestimated in cases where the soft flux includes a significant contribution from star formation, rather than AGN scattered flux. This does not affect the derived maximum AGN luminosity, as this depends only on the reflection normalisation (as well as the assumed solid angle and inclination of the reflector). In model A, we include a reflection component with fixed $R=1$, which should be a reasonable representation for a type 1 (unabsorbed) AGN or a Compton-thin type 2. Model B has no intrinsic absorption, but $A_{\rm ref}$ is free, so the reflection component can dominate, appropriate for Compton thick, type 2 sources. The  {\small XSPEC} representation of Model A is therefore {\small WABS*ZWABS(POWERLAW + PEXMON)}. The  {\small XSPEC} representation of Model B is {\small WABS*(POWERLAW + PEXMON)}. Despite the additional component in model A compared to model B, they have the same number of free parameters because in model B the Compton reflection strength is allowed to vary. To help assess the improvement in the spectral fitting due to the intrinsic absorption (zwabs) or neutral reflection (pexmon) components we also fitted a simple Galactic absorption plus fixed $\Gamma=1.9$ model, {\small WABS*POWERLAW} in {\small XSPEC} (Model C). For each of the three models described above we determine the best fitting parameters to the data, as well at the C-statistic value for the fit. 

The results of the fits for effective photon index and absorbed power-law (Model A) are shown in Table~\ref{fits}, including the C-stat comparison between Model A and Model C. The results of the fits to the reflection model (Model B) are given in Table~\ref{modelB}. A comparison of the C-statistic for the three models is given in Table~\ref{models}, along with the designated best fitting model. We adopt a threshold value of $\Delta C=5$ compared to model C. If the improvement in likelihood exceeds this value we adopted the more complex model (A or B) and choose between A and B based on which delivers the smaller C-statistic. The threshold C-value was determined using Monte Carlo simulations of power law spectra with typical signal-to-noise ratio and background. We performed 10,000 such simulations, fitting them with an absorbed power law (model A) and a reflection model (model B), and comparing the fits to model C . We find that in 1 per cent of the simulations $\Delta C$ exceeded 5.0 (2.4) for model A (model B). The presence of absorption or reflection can therefore be inferred at a minimum of 99 per cent confidence if $\Delta$C exceeds this value. 

The distribution of effective photon index for the SMGs is shown in Fig.~\ref{gamma_hist}, and compared to that from A05b. It can be seen clearly that the SMGs in our sample have, on average, larger values of $\Gamma$ (indicating unobscured X-ray spectra) than those from A05. This is supported by results from the fits to Model A. As can be seen from Table~\ref{fits}, four of the X-ray detected submm galaxies have spectra consistent with large intrinsic absorption ($N_{\rm H}>10^{23}$ cm$^{-2}$: GN04, GN19, GN22 and GN39). A further two sources (GN25 and GN40) show signs of more moderate absorption in their spectra. The majority of the X-ray detected submm galaxies (10/16) are therefore consistent with no intrinsic absorption. The results of the Model A spectral fitting are consistent with the spectral fitting results from A05b for the sources common to both samples.  

All 6 sources that show signs of absorption, and only those 6, also show large improvements (compared to model C) when fitted with model B, the neutral reflection model. Results from fits to this model are given in Table~\ref{modelB}. The reflection fractions range from $\sim 5$ (GN40), to being completely reflection dominated (GN22, GN39-Southern source). Comparing models A and B, (Table~\ref{models}), we find that in GN04 (Northern source), GN19 and GN39 (Northern source) the reflection dominated model gives a lower value of the C-statistic, with the other objects showing a better fit to the absorbed power law. Typically the difference between the two models is not large, however. Exceptions are GN04 (Southern source), GN25, GN39 (Southern source) and GN40 where the transmission model is better (all with $\Delta C>5$ compared to the other model). In GN04 and GN39, which each have two X-ray  counterparts, each counterpart shows significant obscuration. However in each case the Southern source is best fit by the transmission model while the Northern source is marginally better fit by the reflection model. 

Given the limited photon statistics in our spectra, it may not always be possible to identify Compton thick AGN purely on the basis of their X-ray spectral shape. This is especially so given that the X-ray emission can be heavily suppressed by large columns of gas. Thus, even relatively weak X-ray emitters which do not show evidence for absorption or reflection in the spectral fits could still harbour an AGN if it is very heavily obscured. It is possible to estimate the maximum contribution of a Compton-thick AGN to the data using model B, under the assumption that some fraction of the AGN light is viewed via Compton reflection. We adopt the simple assumption that the reflection component is seen without any obscuration, whence the normalization $A_{\rm ref}$ should represent the strength of the illuminating power law, even if that power law is completely obscured from view. As before we assume a slab geometry with an inclination of $60^{\circ}$ and a solid angle $\Omega/2\pi=1$, which are typical parameters, but strongly caution that this introduces uncertainty as the precise obscuring geometry is not known. The spectral fits in model B yield an estimate of $A_{\rm ref}$ and the value of the upper limit  (calculated using $\Delta C=2.7$, corresponding to the 90 per cent confidence limit) in all cases, even if the reflection component does not improve the fit. Table~\ref{modelB} shows the best estimate of the intrinsic AGN luminosity derived using the best estimate of  $A_{\rm ref}$, as long as it exceeds the observed luminosity. We also show the estimated maximum AGN luminosity based on the upper limit to $A_{\rm ref}$, i.e. the luminosity of the illuminating power law which would produce the maximum reflection component tolerated by the data. Again this is quoted only if it exceeds the observed luminosity, otherwise the latter is adopted. 

\begin{table}
\begin{center}
\caption{Results from the spectral fitting using Model B. Errors correspond to the 90 per cent confidence level. 
Col. (1): Submm source name.
Col. (2): Reflection parameter: the ratio of the normalisation parameter
between the reflected component to the direct component.
Col. (3): Improvement in the C-statistic for Model B compared to a $\Gamma$=1.9
power-law spectrum with no intrinsic absorption (Model C).
Col. (4): Directly viewed 2--$10\,$keV luminosity from best fit values to model.
Col. (5): Intrinsic 2--$10\,$keV luminosity of AGN from best fit parameter values
for Model B.
Col. (6): Maximum intrinsic 2--$10\,$keV luminosity of AGN from Model B spectral
fitting. All luminosities are in erg s$^{-1}$.
}
\label{modelB}
\begin{tabular}{@{}lccccc@{}}
\hline
ID    & R       &$\Delta$C-stat & log Lx&log Lx &log Lx  \\
      &         &               & obs. &intr. AGN&max AGN \\
(1)  & (2) & (3) & (4) & (5) & (6) \\     
\hline
GN01  & $<0.8$              & 0.1   & 43.6 & -- & 43.6 \\
GN04$^{a}$ & $>170$     & 113.8 & 42.9 & 44.1 & 44.2  \\
~~~~~~~~$^{b}$  & $17^{+5}_{-6}$       & 31.3  & 43.1 & 44.1 & 44.2  \\
GN06  & $<15.7$     & 0.8  & 42.5 & 43.1 & 43.6  \\
GN10  & $<16.4$      & 0.8   & 42.7 & 43.4 & 43.8  \\
GN12  & $<3.0$                & 0.0  & 42.4  & -- & 42.9  \\
GN13  & $<7.2$                 & 0.0   & 40.9 & -- & 41.8 \\
GN15  & $<2.2$              & 0.0   & 43.0 & -- & 43.4  \\
GN17  & $<6.2$              & 0.0   & 41.9  & -- & 42.7   \\
GN19  & $103^{+21} _{-25}$         & 37.4  & 42.9 & 44.1 & 44.2   \\
GN22  & $>38$         & 12.8  & 42.6 & 43.8 & 44.0 \\
GN23  & $<4.3$           & 0.0   & 42.9 & -- & 43.5 \\
GN25  & $76^{+16}_{-18}$           & 77.6  & 42.7 & 43.8 & 43.9 \\
GN26  & $<5.9$              & 0.0   & 42.1 & -- & 42.9 \\
GN30  & $<10.5$            & 0.0   & 41.8 & -- & 42.8   \\
GN39$^{b}$ & $24^{+13}_{-13}$           & 11.2   & 42.6 & 43.6 & 43.8  \\
~~~~~~~~$^{a}$& $>140$         & 42.8  & 42.5 & 43.7 & 43.9  \\
GN40  & $3.9^{+2.1}_{-1.7}$       & 13.8  & 43.6 & 44.1 & 44.3  \\
\hline
\end{tabular}
\end{center}
$^a$Spectrum of Southern X-ray source.\\
$^b$Spectrum of Northern X-ray source.\\
\end{table}

\begin{table}
\begin{center}
\caption{C-statistic values for models A, B and C.
Model A: wabs*zwabs*(powerlaw+pexmon) with fixed $\Gamma$=1.9 and pexmon
normalisation fixed to powerlaw normalisation.
Model B: wabs*(powerlaw+pexmon) with fixed $\Gamma=1.9$.
Model C: wabs*powerlaw with fixed $\Gamma=1.9$.}
\label{models}
\begin{tabular}{@{}lcccc@{}}
\hline
ID    & C-stat & C-stat&C-stat &Best-fit \\
      & Model A&Model B&Model C&Model\\
\hline
GN01  &    22.4 & 21.7  & 21.8  & C\\ 
GN04$^{a}$ &    33.3 & 41.4  & 155.2 & A\\ 
~~~~~~~~$^{b}$ &    33.2 & 32.6  & 63.9  & B\\ 
GN06  &    40.4 & 39.6  & 40.4  & C\\ 
GN10  &    31.8 & 31.7  & 32.5  & C\\ 
GN12  &    35.1 & 35.0  & 35.1  & C\\ 
GN13  &    30.4 & 30.3  & 30.3  & C\\ 
GN15  &    28.1 & 28.2  & 28.2  & C\\ 
GN17  &    36.8 & 36.8  & 36.8  & C\\ 
GN19  &    32.9 & 30.4  & 67.8  & B\\ 
GN22  &    26.1 & 28.4  & 41.2  & A\\ 
GN23  &    40.4 & 40.3  & 40.3  & C\\ 
GN25  &    39.4 & 44.6  & 122.2 & A\\ 
GN26  &    19.6 & 19.6  & 19.6  & C\\ 
GN30  &    37.7 & 37.6  & 37.6  & C\\ 
GN39$^{b}$ &    43.9 & 41.4  & 52.6  & B\\ 
~~~~~~~$^{a}$ &    30.8 & 39.0  & 81.8  & A\\ 
GN40  &    32.1 & 42.6  & 56.4  & A\\ 
\hline
\end{tabular}
\end{center}
$^a$Spectrum of Southern X-ray source.\\
$^b$Spectrum of Northern X-ray source.\\
\end{table}

\subsection{Stacking of undetected SMGs}

Given the very high X-ray detection rate of the SMGs in the sample, it seems likely that the undetected sources do have significant X-ray emission that is below the detection limit of the CDF-N data. The mean properties of the undetected SMGs can therefore be constrained by stacking, which has proved very useful for other high redshift galaxy populations (e.g.~\citealt{brandt01,horn01,nandra02,laird05,laird06}). 

We use a stacking procedure identical to that describe in \citet{laird05} and \citet{laird06}, which is similar to that used by \citet{nandra02}, and we include only a very brief outline here. Counts at the positions of the IRAC counterparts to the SMGs are extracted using a circular aperture and  summed to find the total counts for the SMGs in the sample. These are then compared to the background counts, which are estimated from a source-masked image using randomly-shuffled positions that are 5--10 arcsec from the SMGs. The background estimate is repeated 1000 times.  We use an aperture radius of 1.25 arcsec to extract the counts from the soft, hard and full bands. This empirically determined value was found to give the strongest signal for a sample of stacked Balmer Break galaxies in the same field \citep{laird05} and the SMGs in this work. 18 SMGs fall within 9 arcmin from the aimpoint of the {\it Chandra\/} observations and we include these in our stacking sample.

\begin{figure}
\begin{center}
\includegraphics[width=125mm,angle=90]{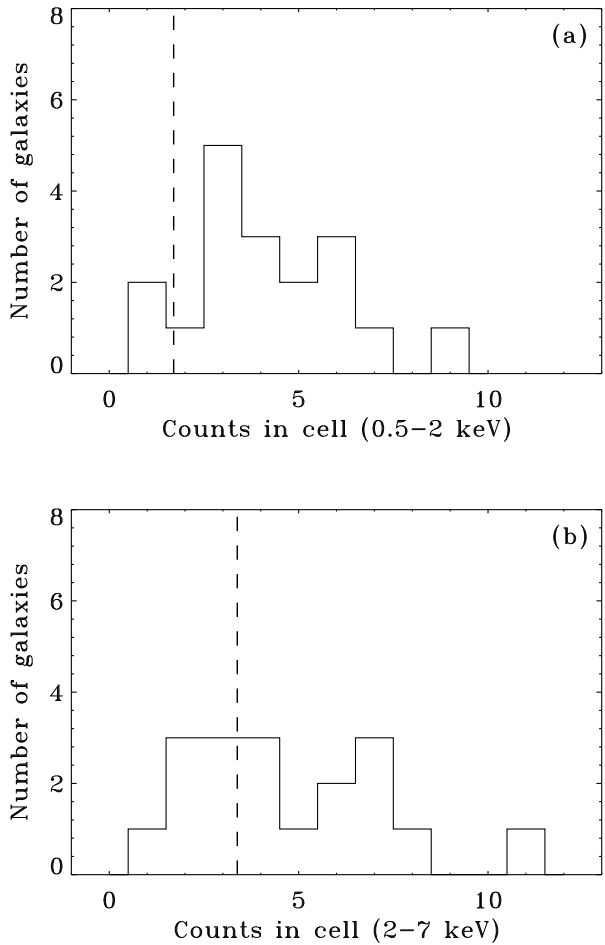}
\caption{
Counts distribution for the undetected SMGs in (a) the 0.5--$2\,$keV and (b) 2--$7\,$keV bands. The vertical dashed line denotes the mean background counts per extraction cell. }
\label{counts_cell}
\end{center}
\end{figure}

The results of the stacking are shown in Table~\ref{stacking}.  The sample of SMGs is significantly detected in the soft, hard and full bands, however we strongly caution that the hard band signal is sensitive to the extraction radius used and the detection significance, $\sigma$, is only greater than 3 for an extraction radius of 1.25 arcsec. Fig.~\ref{counts_cell} shows the counts detected in each extraction cell in the soft and hard bands and demonstrates that the stacked signal is representative of the whole sample and is not dominated by a few sources. The fluxes are calculated assuming a spectrum with $\Gamma=1.9$ and Galactic absorption. The galaxies in the sample have redshifts ranging from 0.8 to 4.1 so to calculate the mean 2--$10\,$keV X-ray luminosity of the sample we use a weighted average method which assumes that each object in the stack is as intrinsically luminous. For every source in the sample we calculate the approximate fraction that it contributes to the flux by weighting by the ratio of the number of counts in the extraction cell for that object to the total number of counts detected for the sample. For each object we are then able to calculate an approximate luminosity. These are then averaged to get the mean L$_{\rmn X}$ (2--$10\,$keV) for the sample. In each band we convert to rest frame 2--$10\,$keV assuming $\Gamma=1.9$. The X-ray luminosity can be used to provide an estimate of the mean star formation rate (SFR) of the sample, under the assumption that the X-ray emission from the SMGs is dominated by star formation processes and not AGNs (e.g. \citealt{grimm03,ranalli03,persic04}).  However it should be noted that the X-ray SFR calibration is subject to considerable uncertainty, and in galaxies with very high SFRs the X-ray emission may be significantly contaminated by an obscured AGN (e.g. \citealt{persic04}). Here we calculate the SFRs using the \citet{ranalli03} relation; using the \citet{persic04} relation, for example, would result in SFRs a factor of 5 lower. 

Using the full band signal, we estimate an average star formation rate of $136\pm24$~M$_{\odot}$ yr$^{-1}$. Interestingly, the soft band stack implies a SFR $\sim 1.5$ times less than the full band, at the 1.5--2$\sigma$ level. This, and the detection of the stack in the hard band at all (which at the mean redshift of 2.1 corresponds to the 6--$20\,$keV rest frame), implies there may be significant contamination of the stacked signal by AGN. Alternatively, in these very gas and dust-rich environments, star forming X-rays could suffer significant obscuration. However, as noted above, the significance of the stacked hard band signal is highly sensitive to the extraction radius used and is only $>3\sigma$ for the radius used here (1.25 arcsec). Therefore we hesitate to draw any robust conclusions from this result.

\begin{table*}
\begin{minipage}{176mm}
\caption{Stacking results of undetected SMGs. All errors are 1$\sigma$ .
Col.(1): Observed-frame energy band. 
Col.(2): Mean redshift. 
Col.(3): Detection significance.
Col.(4): X-ray flux per galaxy. 0.5--$2\,$keV, 2--$10\,$keV and 0.5--$10\,$keV fluxes are given for soft, hard
and full bands, respectively. 
Col.(5): 2--$10\,$keV X-ray luminosity per galaxy.  Luminosities corrected to rest frame 2--$10\,$keV using $\Gamma=1.9$.
Col.(6): X-ray derived mean SFR, using the \citet{ranalli03} relation. Errors are statistical only.}
\label{stacking}
\begin{center} 
\begin{tabular}{@{}lccccc@{}}
\hline
Band & $\langle z \rangle$ & $\sigma$ & F$_{\rmn{X}}$ & L$_{\rmn{X}}$ & SFR$_{\rmn{X}}$ \\
    &      &      & ($10^{-17}~\rmn{erg~cm^{-2}~s^{-1}}$) & ($10^{41}~\rmn{erg~s^{-1}}$) & (${\rm M}_\odot\,{\rm yr}^{-1}$) \\ 
(1) & (2)  & (3)  & (4)  & (5)  & (6)  \\
\hline
Soft & 2.1  & 7.84 &$1.06\pm0.20$ & $4.32\pm0.83$ &   $86\pm17$  \\
Hard & 2.1 & 3.15  &$3.18\pm1.21$ & $8.22\pm3.13$ &   $164\pm62$ \\
Full & 2.1 &  7.10   & $4.31\pm0.78$ &$6.81\pm1.23$ & $136\pm24$ \\
\hline
\end{tabular}
\end{center}
\end{minipage}
\end{table*}

\section{Discussion}

We have presented a detailed analysis of the X-ray properties of submm-selected galaxies in the GOODS-N region. Using the 2-Ms {\it Chandra\/} exposure, and accurate positions for the SMGs from deep radio and {\it Spitzer\/} imaging, we have been able to characterise the X-ray emission of these sources and assess their AGN content. For directly detected sources, we have used the X-ray spectra to determine accurate luminosities and absorption properties. For the undetected sources, we have determined the average properties by stacking. 

\subsection{Origin of the X-ray emission and AGN content of SMGs} 

\begin{figure*}
\begin{center}
\includegraphics[width=55mm,angle=90]{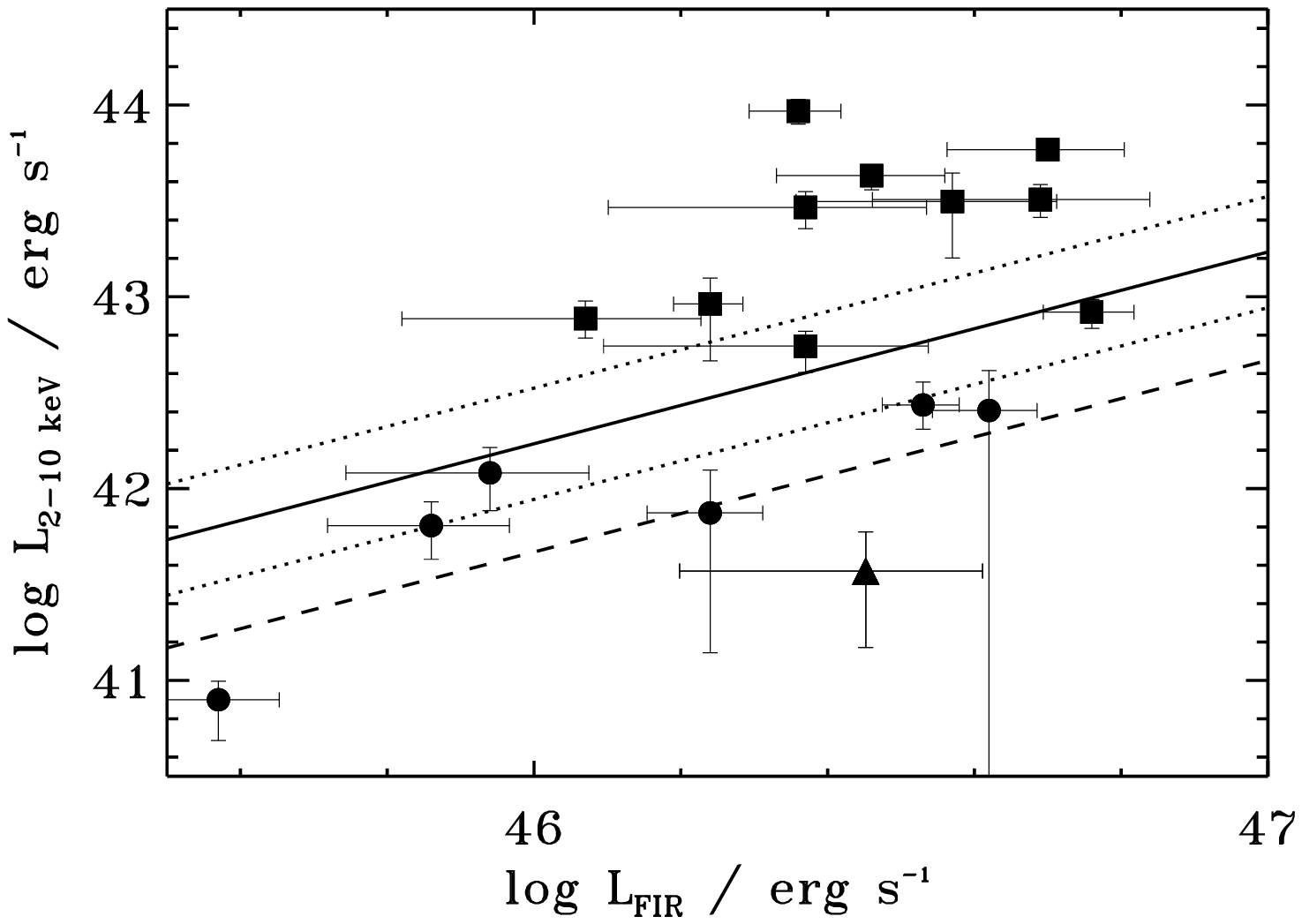}
\hspace{1cm}
\includegraphics[width=55mm,angle=90]{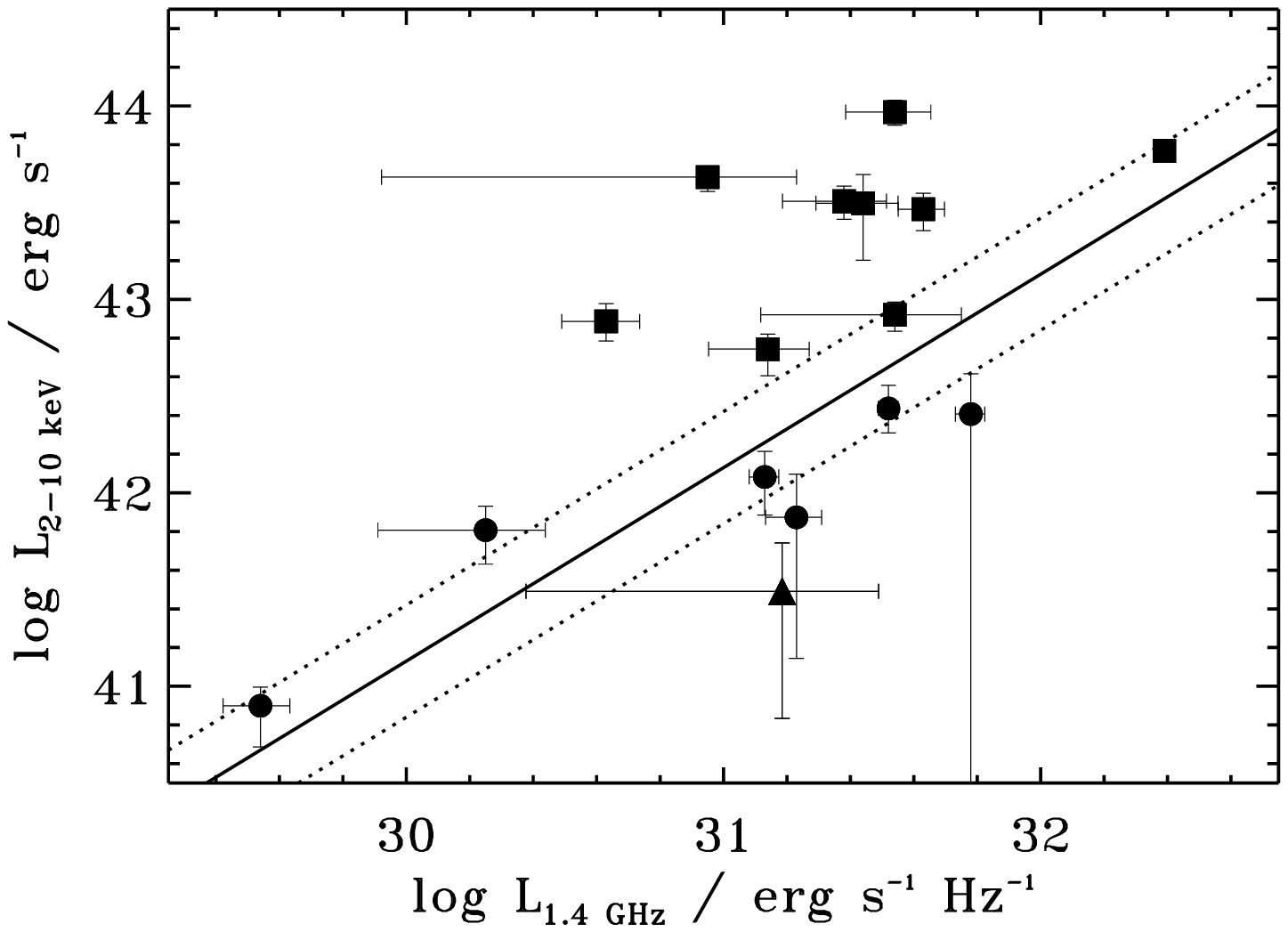}
\caption{Rest-frame 2--$10\,$keV luminosity for the X-ray detected SMGs
against (left) FIR bolometric luminosity and (right)
rest-frame 1.4 GHz radio luminosity. The former are taken from \citet{pope06} and
the latter have been extrapolated from the observed 1.4 GHz fluxes
assuming a power law with $\alpha=-0.75$. The filled circles show the SMGs where the X-ray emission is consistent with that from purely stellar processes. The SMGs where the origin of the X-ray emission is either an AGN or ambiguous are shown as filled squares. 
The X-ray luminosities are those calculated using Model A and have been corrected for intrinsic absorption where relevant (Table~\ref{fits}). 
The filled triangles show the stacking result for the undetected SMGs. The
2--$10\,$keV luminosities have been calculated from the soft band stacking signal, which was the most significant. The right hand plot shows the mean X-ray luminosity for the 13 SMGs in the stack with radio detections. 
The solid and dotted lines in both cases show the local relations for purely star forming galaxies from \citet{ranalli03}. The dashed line in the left plot shows the average X-ray to FIR ratio found for a sample of ultra-luminous infrared galaxies that are not dominated by AGN \citep{teng05}. For GN04 and GN39 the total FIR, radio and X-ray emission is plotted for these sources, as both components are expected to be contributing to the sub-mm flux. 
The X-ray luminosities of most of the SMGs are consistent with this relation, but at high X-ray luminosities ($>10^{43}$~erg s$^{-1}$) there is a clear departure whereby the objects are dominated by an AGN in the X-ray. 
}
\label{lx}
\end{center}
\end{figure*}

We find that the direct X-ray detection rate of submm-selected galaxies in the 2-Ms {\it Chandra\/} data is high, at $46 \pm 8$ per cent.  The detected sources cover a wide range of redshifts, luminosities and X-ray-to-optical flux ratios. The X-ray detected SMGs are thus heterogeneous,  much like the general SCUBA population (e.g.~\citealt{ivison00}). Comparing their X-ray luminosities with the FIR and radio luminosities (Fig.~\ref{lx}), we find that around half of the X-ray detected SMGs are consistent with the X-ray emission arising purely from stellar processes. Using X-ray luminosity as a proxy for SFR (Table~\ref{fits}), the X-ray derived SFRs for these galaxies reach at least $500$~M$_{\odot}$ yr$^{-1}$, and possibly $>$1000~M$_{\odot}$~yr$^{-1}$ in three extreme cases. GN10 (the highest redshift source in the sample) is the most extreme example, with an X-ray luminosity of almost $10^{43}$~erg s$^{-1}$ and star formation rate close of $\sim1700$~M$_{\odot}$ yr$^{-1}$. While extreme, X-ray derived SFRs are subject to considerable uncertainties, especially at large SFRs (e.g. \citealt{persic04,teng05}), and this is fully consistent with the large SFRs calculated from the FIR and polycyclic aromatic hydrocarbon (PAH) luminosities (P06, \citealt{pope08}). 

Seven of the SMGs show X-ray luminosities well in excess of that expected from star formation, however, and these objects almost certainly host AGN. The number of unambiguous AGN in our sample is $20\pm 7$ per cent, rising to $29\pm8$ per cent if we include the three ambiguous cases which apparently have very high star formation rates. The AGN content of these submm-selected sources is certainly very high, but the data do not indicate that strong accretion activity is universal in SMGs. The very high X-ray detection rate of SMGs is likely to be partially due to their very high star formation rates. Furthermore, as we discuss below, at least some SMGs are likely to have their FIR flux powered by an AGN. It is therefore unclear at this point whether the strong overlap between the SMG and X-ray populations points towards a connection between intense star formation and accretion activity in these sources.  

Our results can be compared with those of A05a, who found an even higher AGN fraction than we do, concluding that black holes in SMGs are growing near continuously during periods of intense star formation. The differing conclusions may seem surprising, as we have studied SMGs in the same region (GOODS/CDF-N) as analyzed by A05a, and are using essentially the same X-ray data. Indeed 6 (8) of the 16 X-ray detected SMGs studied here were also in the A05a (A05b) samples (Table~\ref{sample}). For the common sources, our classifcation of the origin of the X-ray emission agrees with A05a and A05b, and thus the main differences can be attributed to the selection and analysis of the submm samples. 

The primary difference is that we are using purely submm-selected sources (predominantly from \citealt{pope05}), whereas A05a use a mixture of  SMGs and optically faint radio sources with followup SCUBA photometry. The AGN fraction of SMGs found in this work ($20-29\pm7$ per cent, depending on the origin of the X-ray emission in the three ambiguous sources in Table~\ref{fits}) is consistent with the lower limit of the AGN fraction quoted by A05a ($>38^{+12}_{-10}$~per cent).
A05a assigned the AGN fraction as a lower limit because they were only able to assess the X-ray properties of SMGs with radio counterparts and spectroscopic redshifts. The SMGs which were radio-undetected in the A05a sample might have been X-ray sources (and hence AGN), but without accurate positions they were unable to tell. As noted by A05a themselves, the radio selection can cause a potential bias in favour of selecting AGN. Requiring a spectroscopic identification can cause a further bias, due to the availability of UV/optical lines in AGN spectra. Using a deeper radio map, and {\it Spitzer\/} data, we have good identifications and accurate positions for the vast majority of our SMGs, and we find that the high AGN fraction among spectroscopically identified, radio-detected submm galaxies fails to extend to the general SMG population. An additional difference worthy of note here is the submm significance threshold. \citet{pope05} applied a strict limit of $3.5\sigma$ to be considered, whereas many of the A05a SMGs have lower significance.

Our excellent positions have enabled us, for the first time, to perform a stacking analysis of the X-ray undetected submm population. We find a very clear soft band detection, and tentative evidence for hard band  emission. The mean X-ray luminosity suggests an average star formation rate of $\sim 150$~M$_{\odot}$ yr$^{-1}$, with the full band delivering a higher star formation rate than the soft X-ray. Including the 9 SMGs from Table~\ref{fits} that are not classified as AGN dominated then the average X-ray derived SFR for the SMGs is $\sim 300$~M$_{\odot}$ yr$^{-1}$. This is considerably less than the mean SFR of $\sim 1200$~M$_{\odot}$ yr$^{-1}$ estimated from the FIR luminosity (P06). With the strong caveats that calibration of X-ray derived SFRs is uncertain (especially at high SFRs),  that the stacked signal may be due to emission from obscured AGN, and that the SFRs from the FIR may be overestimated, there is thus some suggestion that the SMGs may host heavily obscured starbursts, with column densities exceeding $10^{22}$~cm$^{-2}$.

\subsection{Evidence for obscuration}

A05b presented an analysis of the X-ray spectra of the A05a sample, concluding that the majority of X-ray emitting SMGs are heavily obscured, supporting the interpretation of the X-ray emission being from an AGN. For our (small) sample of 7 SMGs that we classify as AGN we confirm this, and X-ray absorption appears to be an effective tool in discriminating between AGN and star forming SMGs. All but one (GN01) of the objects we identify as an AGN shows evidence for obscuration. The column densities in the others are of order $\sim 10^{23}$ cm$^{-2}$. Three sources are candidate Compton thick objects. GN04 (Northern counterpart), GN19 and GN39 (Northern counterpart) have large column densities ($\gtrsim10^{23}$~cm$^{-2}$) and are well fitted by the Compton reflection dominated model but the transmission model provides almost as good a fit. On the other hand, overall the X-ray detected sources in our sample show far less evidence for widespread X-ray absorption than was found by A05b. From the distributions of effective photon index (Fig.~\ref{gamma_hist}) it is clear that SMGs in our sample have, on average, larger values of $\Gamma$ than those in A05b. The K-S probability that the two samples are drawn from the same underlying galaxy distribution is 0.02. Excluding the 8 sources common to both samples, which are selected from the same galaxy population (see Table~\ref{sample} for details), the probability that the two samples are drawn from the same distribution is 0.005. We conclude that the majority ($\sim 60\pm12$ per cent) of X-ray emitting SMGs are unobscured, consistent with the interpretation that the X-ray emission is from stellar processes.  

\subsection{Can AGN activity power the submm emission?}

\begin{figure}
\begin{center}
\includegraphics[width=75mm,angle=90]{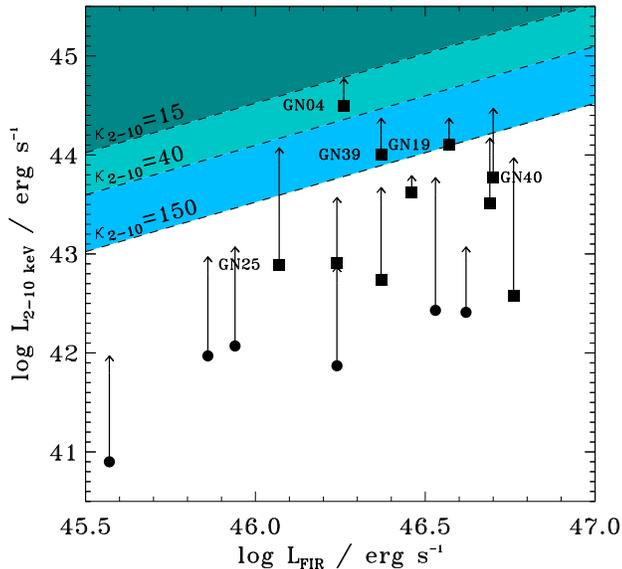}
\caption{Rest-frame 2--$10\,$keV luminosity for the X-ray detected SMGs
against FIR bolometric luminosity. The FIR bolometric luminosities and the symbols are the same as in Fig.~\ref{lx}.
The lower values of the X-ray luminosities are our best estimate of the intrinsic luminosities for each source, calculated for the best-fitting model (Model A, B or C), as given in Table~\ref{models}. The upper values for the X-ray luminosities are the maximum intrinsic values allowed by the reflection model (Model B, Table~\ref{modelB}) and represent the maximum possible AGN contribution.The diagonal lines indicate an AGN-to-FIR bolometric luminosity ratio of 0.5 for 2-10 keV bolometric corrections of $\kappa_{2{-}10}=15, 40$~and $150$. For sources lying in the shaded region above the lines the total bolometric luminosity is AGN dominated ($> 50$ per cent contribution). The five AGN where the bolometric luminosity could be AGN dominated are identified. 
}
\label{bol_lum}
\end{center}
\end{figure}

Our X-ray spectral analysis can be used to estimate the possible contribution of an AGN to the submm emission. For unobscured or Compton thin AGN this is relatively straightforward, albeit subject to substantial uncertainty. In such cases the absorption-corrected X-ray luminosity in, say, the 2--$10\,$keV rest frame band, can be converted using a bolometric correction and compared to the FIR luminosity. The bolometric correction from 2--$10\,$keV has been discussed extensively in the literature (e.g.~\citealt{padovani88}; \citealt{elvis94}; \citealt{barger05}). There is a very wide range of bolometric  corrections, and suggestions that it depends on luminosity (e.g.~\citealt{marconi04}; \citealt{hopkins07}) and/or Eddington ratio \citep{vasudevan07}. The `canonical' value of \citet{elvis94} is $\kappa_{2{-}10}=40$, but the range given by \citet{vasudevan09} covers approximately one order of magnitude $\kappa_{2{-}10}=15$--150. Adopting the \citet{elvis94} correction, we find that the bolometric luminosity is dominated ($>$ 50~per cent contribution) by an AGN in just one of the SMGs, that being GN04. If the upper end of the possible $\kappa_{2{-}10}$ is used then the bolometric luminosities of GN19 and GN39 might also be AGN dominated (Fig.~\ref{bol_lum}). Apparently, AGN make little contribution to the total submm power in the overall population. 

On the other hand, one could hypothesise -- as has been argued by e.g.~A05b -- that many submm sources might harbour very heavily obscured, Compton thick AGN. Using our reflection model fits, it has been possible to estimate the AGN contribution to the luminosity even in sources where we find no evidence for AGN activity. The key point is that a luminous AGN should be revealed via a Compton scattered X-ray continuum, even when the direct continuum is obscured from view. Such a Compton reflection component, with accompanying intense iron K$\alpha$ line emission, is a defining characteristic of Compton thick AGN in the local Universe (e.g.~\citealt{reynolds94}; \citealt{matt00}). The lack of such a component in the majority of spectra in our sample limits the possible contribution of an AGN to the submm typically to between 1--10~per cent, assuming the \citet{elvis94} bolometric correction. The exceptions are the three objects noted above. Taking the estimates of the maximum possible contribution allowed by the spectral fits, and the most extreme bolometric correction, we find that GN25 and GN40 might also have as much as 50~per cent of their submm emission powered by the AGN (Fig.~\ref{bol_lum}). For Compton reflection dominated objects, such estimates are very uncertain, as the geometry of the reflecting medium (including the degree of self-obscuration) is not well known, and this will strongly affect the strength of the reflection signatures. Overall, however, we can conclude reasonably robustly that around 85~per cent of SMGs have their FIR emission dominated by star formation, whether or not an AGN is present. 

An alternative way of diagnosing the presence of an AGN and its contribution to the FIR emission is via mid-IR spectroscopy. \citet{valiante07}, \citet{pope08} and \citet{menendez09} have analysed IR spectroscopy of more than 40 SMGs, a few of which overlap with the  sample analysed here. These works showed that the bolometric luminosity of most SMGs is starburst dominated and that mid-IR dominant AGN in SMGs are rarely found: only 15--17 per cent of SMGs are dominated by an AGN in the mid-IR \citep{pope08,menendez09}.  For the sources common to this sample, the mid-IR spectral diagnostics generally agree with the X-ray diagnostics. The exceptions being GN19 and GN39 which are both clear AGN from the X-ray but are starburst dominated based on their mid-IR spectral properties \citep{pope08}. 

\section{Summary}

We have presented an estimate of the AGN fraction and X-ray spectral properties of submm-selected galaxies in the GOODS-N field, based on the deep (2--Ms) {\it Chandra\/} observation. While SMGs show a high X-ray detection fraction ($45\pm8$~per cent), and also AGN fraction ($20-29\pm7$~per cent), the latter is more modest compared to some previous estimates. Most X-ray detected SMGs have low X-ray luminosity and are unobscured, so are probably dominated by stellar emission in the X-ray. On the other hand, the (minority) AGN SMGs are obscured with moderate column densities, and we find three candidate Compton thick objects. We have estimated the AGN contribution to the bolometric luminosity and find only three objects in our sample in which accretion is the dominant power source. The X-ray spectra therefore indicate that star formation dominates the FIR emission in at least $\sim$85~per cent of the submm population, in agreement with mid-IR spectral diagnostics. Overall, we conclude that while intense, coeval star formation and intense accretion is undoubtedly present in some SMGs, this behaviour is far from universal, with only around 10--15~per cent of the population showing clear evidence for this mode of activity.  

\section*{Acknowledgements}
We thank the anonymous referee for careful reading of the manuscript that has improved the quality of this paper. 
ESL acknowledges the support of STFC. KN acknowledges the Leverhulme trust.
AP acknowledges support provided by NASA through the {\it Spitzer Space Telescope} Fellowship Program, through a contract issued by the Jet Propulsion Laboratory, California Institute of Technology under contract with NASA. DS acknowledges support from the Natural Sciences and Engineering Research Council of Canada.

\label{lastpage}

\end{document}